\documentclass[12pt,a4paper]{article}
\usepackage{amsmath,amssymb}
\usepackage{graphicx}
\usepackage[sort]{cite}
\newcommand{\DP}{\Delta\Pi}
\newcommand{\APF}[1]{A_{0}^{(#1)}}
\newcommand{\KG}[2]{K_{#1}^{(#2)}}
\newcommand{\KGt}[2]{\tilde{K}_{#1}^{(#2)}}
\newcommand{\mmu}{m_{\mu}}
\newcommand{\iep}{i\varepsilon}
\newcommand{\Li}[1]{{\rm Li}_{#1}}

\allowdisplaybreaks

\begin{document}

\begin{center}

{\Large\bf
Spacelike and timelike kernels for the hadronic vacuum polarization
contribution to the muon lifetime

}

\vskip10mm

{\large A.V.~Nesterenko}

\vskip7.5mm

{\small\it Bogoliubov Laboratory of Theoretical Physics,
Joint Institute for Nuclear Research,\\
Dubna, 141980, Russian Federation}

\end{center}

\vskip5mm

\noindent
\centerline{\bf Abstract}

\vskip2.5mm

\centerline{\parbox[t]{150mm}{%
The explicit expressions for the spacelike kernels~$\KG{\Pi}{2}(Q^2)$
and~$\KG{D}{2}(Q^2)$ for the hadronic vacuum polarization contribution to
the muon lifetime~$\Delta\Gamma_{\text{had}}$ are derived. The obtained
results enable one to evaluate~$\Delta\Gamma_{\text{had}}$ within such
methods as lattice studies, MUonE project, and others.}}

\vskip12mm

\section{Introduction}
\label{Sect:Intro}

The hadronic vacuum polarization (HVP) contributions play a pivotal role in
two distinct facets of muon physics: muon anomalous magnetic moment
\mbox{$a_{\mu} = (g_{\mu}-2)/2$} and muon lifetime~$\tau_{\mu}$.
Specifically, while the experimental measurement of~$a_{\mu}$ has reached an
exceptional level of accuracy~\cite{Muong-2:2006rrc, Muong-2:2025xyk},
currently its consistency with theoretical evaluation of~$a_{\mu}$ (and,
hence, potential evidence for the existence of a new fundamental physics
beyond the Standard Model) cannot be definitively determined due to
significant tension between the values of
HVP~contribution~$a^{\text{HVP}}_{\mu}$ obtained within spacelike lattice and
timelike data-driven methods and amongst those of the timelike data-driven
method itself~\cite{Aoyama:2020ynm, Aliberti:2025beg}. At~the same time,
rigorous evaluation of HVP contribution~$\Delta\Gamma_{\text{had}}$
to~$\tau_{\mu}$ is essential for precise extraction of a~fundamental
parameter of the electroweak sector of the Standard Model, namely, the Fermi
coupling constant~$G_F$, from the muon lifetime
measurements~\cite{FAST:2007rsc, MuLan:2010shf}.

At present, for HVP~contribution to the muon
lifetime~$\Delta\Gamma_{\text{had}}$ only the timelike kernel ${K}_{R}(s)$ is
available~\cite{vanRitbergen:1998hn}, that restricts its assessment to the
timelike data-driven method. The~primary objective of this study is to obtain
the corresponding spacelike kernels~$K_{\Pi}(Q^2)$ and~$K_{D}(Q^2)$, that
would enable one to evaluate~$\Delta\Gamma_{\text{had}}$ within such
spacelike methods as lattice simulations (see~e.g.,
reviews~\cite{Meyer:2018til, Wittig:2023pcl}) and a highly anticipated MUonE
project~\cite{CarloniCalame:2015obs, MUonE:2016hru}, thereby providing a
comprehensive determination of the Fermi coupling constant~$G_F$.

\section{Methods}
\label{Sect:Methods}

\subsection{Dispersion relations}
\label{Sect:DR}

The theoretical studies of hadronic contributions to a variety of the
Standard Model observables inherently rely on HVP~function~$\Pi(q^2)$, which
is defined as the scalar part of HVP~tensor
\begin{align}
\label{P_Def}
\Pi_{\mu\nu}(q^2) & = i\!\int\!d^4x\,e^{i q x} \bigl\langle 0 \bigl|\,
T\bigl\{J_{\mu}(x)\, J_{\nu}(0)\bigr\} \bigr| 0 \bigr\rangle
= \frac{i}{12\pi^2} (q_{\mu}q_{\nu} - g_{\mu\nu}q^2) \Pi(q^2).
\end{align}
As argued, for example, in Ref.~\cite{Feynman:2019rot}, the
function~$\Pi(q^2)$~(\ref{P_Def}) has a single cut along the positive
semiaxis of real~$q^2$ starting at the hadronic threshold~\mbox{$q^2 \ge
s_{\text{th}}$}. While the once--subtracted Cauchy's integral formula
represents~$\Pi(q^2)$ in terms of its discontinuity across the physical cut
[which is commonly identified with the \mbox{$R$-ratio} of electron-positron
annihilation into hadrons $R(s)$, with \mbox{$s=q^2>0$} being the timelike
kinematic variable], for practical purposes it~is particularly convenient to
employ the logarithmic derivative of~HVP function, namely, the Adler
function~$D(Q^2)$, with~\mbox{$Q^2=-q^2>0$} being the spacelike kinematic
variable. The~complete set of relations expressing the functions~$\Pi(q^2)$,
$D(Q^2)$,~$R(s)$ in terms of one another reads~\cite{Feynman:2019rot,
Adler:1974gd, Radyushkin:1982kg, Pennington:1981cw, Pivovarov:1991bi}
\begin{subequations}
\label{DispRels}
\begin{align}
\label{DispRelsP}
\DP(q^{2},\, q_{0}^{2}) & = (q^2 - q_0^2) \int_{s_{\text{th}}}^{\infty}
\frac{R(\sigma)}{(\sigma-q^2)(\sigma-q_0^2)}\, d\sigma =
- \int_{-q_0^2}^{-q^2} \!D(\sigma)\, \frac{d\sigma}{\sigma},
\\
\label{DispRelsD}
D(Q^2) & = - \frac{d\,\Pi(-Q^2)}{d\,\ln Q^2} =
Q^2\! \int_{s_{\text{th}}}^{\infty} \frac{R(\sigma)}{(\sigma+Q^2)^2}\, d\sigma,
\\
\label{DispRelsR}
R(s) & = \frac{1}{2 \pi i} \lim_{\varepsilon \to 0_{+}}
\Bigl[\Pi(s+i\varepsilon)-\Pi(s-i\varepsilon)\Bigr]\! =
\frac{1}{2 \pi i} \lim_{\varepsilon \to 0_{+}}
\int_{s+i\varepsilon}^{s-i\varepsilon} D(-\zeta)
\frac{d\zeta}{\zeta},
\end{align}
\end{subequations}
where $\DP(q^{2},\, q_{0}^{2}) = \Pi(q^2) - \Pi(q_0^2)$. A~detailed overview
of this matter can be found in, e.g., Chap.~1 of
Ref.~\cite{Nesterenko:2016pmx} and references therein.

\subsection{Hadronic vacuum polarization contribution to~$\tau_{\mu}$}
\label{Sect:TauMuHVP}

The inverse lifetime of the muon can be represented~as
\begin{equation}
\frac{1}{\tau_{\mu}} =
\Gamma\bigl(\mu^{-} \to e^{-}\bar\nu_{e}\nu_{\mu}\bigr) =
\Gamma_{0} + \Delta\Gamma,
\quad
\Gamma_{0} = \frac{G_{F}^{2}m_{\mu}^{5}}{192\pi^{3}},
\end{equation}
where~$m_{\mu}$ is the muon mass and~$\Delta\Gamma$ denotes the higher-order
corrections. The hadronic part of the latter can be cast in terms of either
of the functions~(\ref{DispRels}). Namely, in the \mbox{$\ell$-th}~order in
the electromagnetic coupling
\begin{subequations}
\label{DGhad}
\begin{align}
\label{DGhadP}
\Delta\Gamma_{\text{had}}^{(\ell)} & = \Gamma_{0}\APF{\ell}\!\int_{0}^{\infty}
\KG{\Pi}{\ell}(Q^2) \bar\Pi(Q^2) \frac{d Q^2}{4\mmu^2} =
\Gamma_{0}\APF{\ell}\!\int_{0}^{\infty}
\KGt{\Pi}{\ell}(\zeta) \bar\Pi(4\zeta\mmu^2) d \zeta
\\
\label{DGhadD}
& = \Gamma_{0}\APF{\ell}\!\int_{0}^{\infty}
\KG{D}{\ell}(Q^2) D(Q^2) \frac{d Q^2}{4\mmu^2} =
\Gamma_{0}\APF{\ell}\!\int_{0}^{\infty}
\KGt{D}{\ell}(\zeta) D(4\zeta\mmu^2) d \zeta
\\
\label{DGhadR}
& = \Gamma_{0}\APF{\ell}\!\int_{s_{\text{th}}}^{\infty}
\KG{R}{\ell}(s) R(s) \frac{d s}{4\mmu^2} =
\Gamma_{0}\APF{\ell}\!\int_{\chi}^{\infty}
\KGt{R}{\ell}(\eta) R(4\eta\mmu^2) d \eta,
\qquad
\APF{2} = \frac{1}{3} \biggl(\frac{\alpha}{\pi}\biggr)^{2},
\end{align}
\end{subequations}
where~\mbox{$\bar\Pi(Q^2) = \DP(0,-Q^2) = -\Pi(-Q^2)$} is the subtracted at
zero HVP~function [with
\linebreak
\mbox{$\Pi(0)=0$} being assumed], $\APF{\ell}$ stands
for a constant prefactor, \mbox{$\zeta = Q^2/(4\mmu^2)$} and~\mbox{$\eta =
s/(4\mmu^2)$} are the dimensionless kinematic variables,
\mbox{$\chi=s_{\text{th}}/(4\mmu^2)$}, and~$\KG{\Pi}{\ell}(Q^2)$,
$\KG{D}{\ell}(Q^2)$, $\KG{R}{\ell}(s)$ denote the corresponding kernels.

As noted earlier, only the leading-order (i.e., second order in~the
electromagnetic coupling) timelike kernel~$\KG{R}{2}(s)$ was
calculated~\cite{vanRitbergen:1998hn} [see Eq.~(\ref{KRexpl}) below], whereas
its spacelike counterparts~$\KG{\Pi}{2}(Q^2)$,~$\KG{D}{2}(Q^2)$ and the
higher-order kernels are unavailable at~present. Nonetheless, since the
functions~$\bar\Pi(Q^2)$, $D(Q^2)$, $R(s)$ entering the
representations~(\ref{DGhad}) are interrelated~(\ref{DispRels}), the
corresponding kernels can also be expressed in terms of one another, thereby
enabling one to directly derive the unknown kernels from a known one, see
Refs.~\cite{Nesterenko:2021byp, Balzani:2021del, Aliberti:2025beg} for the
details. Specifically, similarly to the timelike kernels
for~$a^{\text{HVP}}_{\mu}$ (see, for~example, discussion in
Ref.~\cite{Barbieri:1974nc}), the function~$\KG{R}{2}(s)$~(\ref{KRexpl}) has
a single cut along the negative semiaxis of real~$q^2$ starting at the
origin~$q^2 \le 0$, thus making the complete set of relations between the
spacelike and timelike kernels for~$a^{\text{HVP}}_{\mu}$ obtained in
Ref.~\cite{Nesterenko:2021byp} applicable to the current study:
\begin{subequations}
\label{KRels}
\begin{align}
\label{KRelsP}
\KG{\Pi}{\ell}(Q^2) & = \frac{1}{2 \pi i} \lim_{\varepsilon \to 0_{+}}
\Bigl[ \KG{R}{\ell}(-Q^2+\iep) - \KG{R}{\ell}(-Q^2-\iep)\Bigr] =
- \Biggl[\KG{D}{\ell}(Q^2)
+ \frac{d\,\KG{D}{\ell}(Q^2)}{d\,\ln Q^2}\Biggr],
\\
\label{KRelsD}
\KG{D}{\ell}(Q^2) & = \frac{1}{Q^2}
\int_{Q^2}^{\infty} \KG{\Pi}{\ell}(p^{2})\, d p^{2}
=
- \frac{1}{2 \pi i} \lim_{\varepsilon \to 0_{+}}
\frac{1}{Q^2}
\int_{Q^2 + \iep}^{Q^2 - \iep}
\KG{R}{\ell}(-p^2) d p^2,
\\[1.5mm]
\label{KRelsR}
\KG{R}{\ell}(s) & = \frac{1}{s} \int_{0}^{\infty}
\KG{\Pi}{\ell}(Q^2) \frac{Q^2}{s + Q^2}\, d Q^2 =
\int_{0}^{\infty}
\KG{D}{\ell}(Q^2) \frac{Q^2}{(s + Q^2)^2}\, d Q^2.
\end{align}
\end{subequations}
It is worthwhile to note here that the first of the relations~(\ref{KRelsP})
has also been independently derived in a different way
in~Ref.~\cite{Balzani:2021del}.

\section{Results}
\label{Sect:Results}

\subsection{Spacelike kernels for HVP contribution to the muon lifetime}

\begin{figure}[t]
\centerline{\includegraphics[width=100mm,clip]{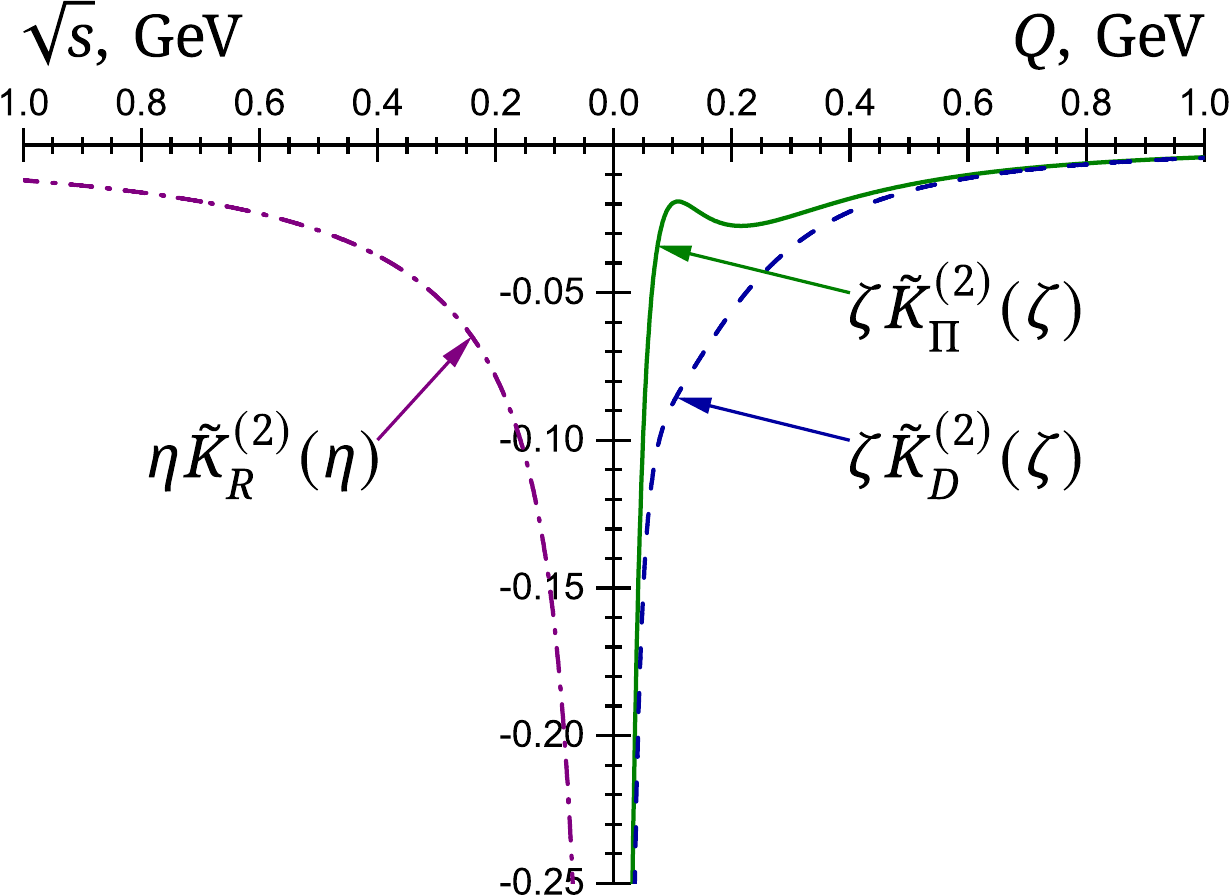}}
\caption{Kernels for the leading-order HVP~contribution to the muon
lifetime~(\ref{DGhad}) in spacelike [\mbox{$Q^2 = -q^2 \ge 0$}, \mbox{$\zeta
= Q^2/(4\mmu^2)$}] and timelike [\mbox{$s = q^2 \ge 0$}, $\eta =
s/(4\mmu^2)$] domains: $\zeta\KGt{\Pi}{2}(\zeta)$ [Eq.~(\ref{KPexpl}), solid
curve], $\zeta \KGt{D}{2}(\zeta)$ [Eq.~(\ref{KDexpl}), dashed curve], and
$\eta \KGt{R}{2}(\eta)$ [Eq.~(\ref{KRexpl}), dot-dashed curve].}
\label{Plot:Kernels}
\end{figure}

Thus, for the leading-order spacelike kernel~$\KG{\Pi}{2}(Q^2)$ for HVP
contribution to the muon lifetime~(\ref{DGhadP}), the first of the
relations~(\ref{KRelsP}) together with Eq.~(\ref{KRexpl}) yields
\begin{align}
\label{KPexpl}
\zeta\KGt{\Pi}{2}(\zeta) & =
\frac{32}{3}\zeta^3 - 10\zeta^2 -\frac{86}{9}\zeta + \frac{5}{3}
+\frac{1}{\psi(-\zeta)}\biggl( \frac{32}{3}\zeta^3 + \frac{22}{3}\zeta^2
-\frac{31}{3}\zeta - \frac{11}{2} \biggr)
\nonumber \\ &
-\biggl( \frac{128}{3}\zeta^4 -48\zeta^2 - \frac{32}{3}\zeta + 1 \biggr)
\ln\biggl[\frac{1+\psi(-\zeta)}{2}\biggr]
\nonumber \\ &
-\zeta^2\biggl( \frac{128}{3}\zeta^2 -\frac{64}{3}\zeta - 32 \biggr)
\psi(-\zeta)\ln\biggl[\frac{1+\psi(-\zeta)}{2}\biggr],
\qquad
\psi(-\zeta) = \sqrt{1+\frac{1}{\zeta}}.
\end{align}
In turn, for the leading-order spacelike kernel~$\KG{D}{2}(Q^2)$ for HVP
contribution to the muon lifetime~(\ref{DGhadD}), the first of the
relations~(\ref{KRelsD}) together with Eq.~(\ref{KPexpl}) leads~to
\begin{align}
\label{KDexpl}
\zeta\KGt{D}{2}(\zeta) & =
-\ln^2\bigl[\phi(\zeta)\bigr]
+\ln\bigl[\phi(\zeta)\bigr]\frac{g_{1}(\zeta)\psi(-\zeta)+g_{2}(\zeta)}
{g_{3}(\zeta)\psi(-\zeta)+g_{4}(\zeta)}
\nonumber \\[1.5mm] &
-2\Li{2}\bigl[1-\phi(\zeta)\bigr]
+\frac{g_{5}(\zeta)\psi(-\zeta)+g_{6}(\zeta)}
{g_{7}(\zeta)\psi(-\zeta)+g_{8}(\zeta)},
\end{align}
where~$\phi(\zeta)=2\zeta\bigl[\psi(-\zeta)-1\bigr]$,
\begin{align}
g_{1}(\zeta) & =\zeta\bigl(-992\zeta^3-1680\zeta^2-796\zeta-86\bigr),
\nonumber \\[1.5mm]
g_{2}(\zeta) & =992\zeta^4+2176\zeta^3+1512\zeta^2+336\zeta+10,
\nonumber \\[1.5mm]
g_{3}(\zeta) & =\zeta\bigl(384\zeta^3+576\zeta^2+240\zeta+24\bigr),
\nonumber \\[1.5mm]
g_{4}(\zeta) & =-\bigl(384\zeta^4+768\zeta^3+480\zeta^2+96\zeta+3\bigr),
\nonumber \\[1.5mm]
g_{5}(\zeta) & =\zeta\bigl(27648\zeta^4+25600\zeta^3-15008\zeta^2-15552\zeta-1842\bigr),
\nonumber \\[1.5mm]
g_{6}(\zeta) & =-\bigl(27648\zeta^5+39424\zeta^4-5664\zeta^3-24528\zeta^2-7222\zeta-171\bigr),
\nonumber \\[1.5mm]
g_{7}(\zeta) & =\zeta\bigl(4608\zeta^2+4608\zeta+864\bigr),
\nonumber \\[1.5mm]
g_{8}(\zeta) & =-\bigl(4608\zeta^3+6912\zeta^2+2592\zeta+144\bigr),
\end{align}
and
\begin{equation}
\label{Li2Def}
\Li{2}(x) = -\int_{0}^{x}\ln(1-t)\frac{d\,t}{t}
\end{equation}
denotes the dilogarithm function.

The spacelike $\zeta\KGt{\Pi}{2}(\zeta)$ [Eq.~(\ref{KPexpl}), solid curve],
$\zeta \KGt{D}{2}(\zeta)$ [Eq.~(\ref{KDexpl}), dashed curve], and timelike
$\eta \KGt{R}{2}(\eta)$ [Eq.~(\ref{KRexpl}), dot-dashed curve] kernels for
the leading-order HVP~contribution to the muon lifetime~(\ref{DGhad}) are
displayed in Fig.~\ref{Plot:Kernels}. Similarly to the timelike
kernel~(\ref{KRexpl}), the spacelike kernels~(\ref{KPexpl})
and~(\ref{KDexpl}) are enhanced at low energies and suppressed at high
energies, namely
\begin{align}
\label{KPexpIR}
\zeta\KGt{\Pi}{2}(\zeta) & \simeq
\frac{5}{3} +\frac{1}{2}\ln(4\zeta) -\frac{13}{2}\sqrt{\zeta}
+\mathcal{O}(\zeta),
\quad
\zeta \to 0_{+},
\\[1.5mm]
\label{KPexpUV}
\zeta\KGt{\Pi}{2}(\zeta) & \simeq
-\frac{1}{10\zeta} + \frac{97}{576\zeta^{2}}
+\mathcal{O}\biggl(\frac{1}{\zeta^3}\biggr),
\quad
\zeta \to \infty,
\\[1.5mm]
\label{KDexpIR}
\zeta\KGt{D}{2}(\zeta) & \simeq
-\frac{19}{16} -\frac{\pi^{2}}{3} -\frac{5}{3}\ln(4\zeta)
-\frac{1}{4}\ln^{2}(4\zeta)
+13\sqrt{z} +\mathcal{O}(\zeta),
\quad
\zeta \to 0_{+},
\\[1.5mm]
\label{KDexpUV}
\zeta\KGt{D}{2}(\zeta) & \simeq
-\frac{1}{10\zeta} + \frac{97}{1152\zeta^{2}}
+\mathcal{O}\biggl(\frac{1}{\zeta^3}\biggr),
\quad
\zeta \to \infty.
\end{align}

\section{Conclusions}
\label{Sect:Concl}

The explicit expressions for the spacelike
kernels~$\KG{\Pi}{2}(Q^2)$~(\ref{KPexpl}) and~$\KG{D}{2}(Q^2)$~(\ref{KDexpl})
for HVP~contribution to the muon lifetime~$\Delta\Gamma_{\text{had}}$ are
derived. The obtained results enable one to
evaluate~$\Delta\Gamma_{\text{had}}$ within such spacelike methods as lattice
studies, MUonE project, and others.

\appendix
\section{Timelike kernel for HVP contribution to the muon lifetime}

The leading-order timelike kernel appearing in Eq.~(\ref{DGhadR}) has been
known since the late 1990s~\cite{vanRitbergen:1998hn}, namely
\begin{align}
\label{KRexpl}
\eta\KGt{R}{2}(\eta) & =
\frac{64}{3}\eta^3 +\frac{20}{3}\eta^2 -\frac{635}{27}\eta -\frac{19}{16}
+ \biggl( \frac{32}{3}\eta^3 +10\eta^2 -\frac{86}{9}\eta -\frac{5}{3} \biggr)\ln(4\eta)
\nonumber \\ &
+ \biggl( \frac{32}{3}\eta^3 -\frac{22}{3}\eta^2 -\frac{31}{3}\eta +\frac{11}{2} \biggr)
\Bigl\{ \ln\bigl[1+\psi(\eta)\bigr] -\ln\bigl[1-\psi(\eta)\bigr] \Bigr\}\frac{1}{\psi(\eta)}
\nonumber \\ &
-\biggl( \frac{128}{3}\eta^4 -48\eta^2 +\frac{32}{3}\eta +1 \biggr)
\biggl\{ \frac{\pi^2}{6}
+\ln\biggl[ \frac{1-\psi(\eta)}{2} \biggr]
 \ln\biggl[ \frac{1+\psi(\eta)}{2} \biggr] \biggr\}
\nonumber \\ &
+ \eta^{2}\biggl( \frac{128}{3}\eta^2 +\frac{64}{3}\eta -32 \biggr)\psi(\eta)
\biggl\{ \Li{2}\biggl[ \frac{1+\psi(\eta)}{2} \biggr]
        -\Li{2}\biggl[ \frac{1-\psi(\eta)}{2} \biggr] \biggr\},
\end{align}
where~$\psi(\eta) = \sqrt{1-\eta^{-1}}$. The kernel~(\ref{KRexpl}) is
enhanced at low energies and suppressed at high energies, specifically
\begin{align}
\label{KRexpIR}
\eta\KGt{R}{2}(\eta) & \simeq
-\frac{19}{16} -\frac{5\pi^2}{12} -\frac{5}{3}\ln(4\eta)
-\frac{1}{4}\ln^{2}(4\eta) +\frac{13}{2}\pi\sqrt{\eta}
+\mathcal{O}(\eta),
\quad
\eta \to 0_{+},
\\[1.5mm]
\label{KRexpUV}
\eta\KGt{R}{2}(\eta) & \simeq
\frac{1}{10\eta}\biggl[\frac{41}{20} - \ln(4\eta)\biggr]
+\frac{1}{576\eta^2}\biggl[\frac{4321}{12} -20\pi^{2} -97\ln(4\eta)\biggr]
+\mathcal{O}\biggl(\frac{1}{\eta^3}\biggr),
\quad
\eta \to \infty.
\end{align}

\bibliography{paper}

\end{document}